\documentclass[12pt]{iopart}
\usepackage{iopams}  
\usepackage{graphicx,epsfig,color}
\usepackage{hyperref}

\newcommand{\beq}{\begin{equation}}
\newcommand{\eeq}{\end{equation}}
\newcommand{\ec}{\end{center}}
\newcommand{\bc}{\begin{center}}
\newcommand{\eea}{\end{eqnarray}}
\newcommand{\bea}{\begin{eqnarray}}
\newcommand{\eeas}{\end{eqnarray*}}
\newcommand{\beas}{\begin{eqnarray*}}
\newcommand{\bd}{\begin{description}}
\newcommand{\ed}{\end{description}}
\newcommand{\mb}{\mathbf}
\newcommand{\vx}{\mathbf{x}}

\newcommand{\ei}{\end{itemize}}
\newcommand{\vk}{\mathbf{k}}

\newcommand{\vq}{\mathbf{q}}

\newcommand{\vj}{\mathbf{j}}

\newcommand{\mysubsection}[2][]

\begin{document}
\title{A Renormalization Scheme and Skewness of Height Fluctuations in $(1+1)$-dimensional VLDS Dynamics}
\author{Tapas Singha \&  Malay K. Nandy}
\address{Department of Physics, Indian Institute of Technology Guwahati, Guwahati 781039, India.}
\ead{s.tapas@iitg.ernet.in \& mknandy@iitg.ernet.in}
\vspace{10pt}
\date{(November 15, 2015)}

\begin{abstract}
 We study the $(1+1)$-dimensional Villain, Lai, and Das Sarma (VLDS) equation driven by a Gaussian 
 white noise and implement a renormalization scheme without rescaling at one-loop order. Using a 
 diagrammatic method, we calculate the renormalized second and third moments in the large-scale and 
 long-time limits. The ensuing skewness value is $S=-0.0441$. This (negative) value is consistent
 with the numerical prediction of Das Sarma \emph{et al.} [Phys. Rev. E {\bf 53} 359 (1996)].
\end{abstract}
\hspace{4pc}

\section{Introduction}
In the last few decades, one of the most widespread, useful and fascinating
topics of research in nonequilibrium statistical physics  is kinetic interface
roughening \cite{book_stanley,krug97,Halpin95,Meakin93,Family_Physica,Family_Physica_1990,
Markov_Epitaxy}. A great deal of experiments and numerical simulations have been carried out in
this field and a few analytical methods have been employed
including the renormalization group (RG) to understand the scaling behavior, exponents and
thereby universality of a growing surface. A linear continuum dynamics of
surface growth is described by the Edwards-Wilkinson \cite{EW} (EW) equation
where the surface profile evolves due to random deposition and surface
tension. 

A nonlinear paradigmatic equation for surface growth, as proposed by
Kardar, Parisi and Zhang (KPZ) \cite{KPZ86,KPZ89}, describes a wide range of surface growth 
phenomena such as the eden model \cite{eden_plischke_85}, ballistic deposition \cite{Meakin86}, restricted 
solid-on-solid model (RSOS), all belonging to a common universality class as they exhibit the same scaling exponents. 

The growth of a thin film using the molecular beam epitaxy (MBE) is
dominated by surface  diffusion \cite{SP92_PRL.68.2035} at high temperatures,
 where atomic evaporation \cite{PlRaLi_87_PRB.35.3485}, desorption,
bulk defects \cite{H92_PRL.68.3048}, hangs and overhangs are negligibly small.
In MBE, atoms are deposited one-by-one for the preparation of high quality thin films \cite{FDA_04}.
On the other hand, at low temperatures, a growing surface would be amorphous instead of
crystalline \cite{Villain_j_Phys_91} because deposited particles would not have
much energy to overcome the height barriers so that they can settle down at the lowest energy positions. 
According to Mullins's theory \cite{Mullins_59} such surface growth obeys the equation of continuity
\beq
\frac{\partial{h}}{\partial t}=-\nabla \cdot \vj
\label{Mullins}
\eeq
where the particle current is $ \vj= \nabla \mu(\vx,t)$ and local chemical
potential $\mu(\vx,t)$ is related to the surface curvature as $\mu(\vx,t)=\kappa \nabla^2 h(\vx,t)$. 
Fourier transforming and rescaling of equation \ref{Mullins} yield roughness, dynamic and growth exponents as 
$\chi=(4-d)/2$, $z=4$ and $\beta=(4-d)/8$ respectively, $d$ being the substrate 
dimension. Wolf and Villain \cite{Wolf_1990} (WV) studied a model with random deposition
and surface diffusion in one dimension, where the particles are relaxed to the minimum energy
positions yielding roughness and dynamic exponents,
$\chi=1.4\pm0.1$ and $z=3.8\pm0.5$, that indicates a different
universality class from those of EW and KPZ. These exponents are close to those obtained
from equation \ref{Mullins}. Das Sarma and Tamborenea (DT)
\cite{sarma-tamborenea_91} investigated a deposition model with surface
relaxation and estimated roughness and  growth exponents as $\chi\approx 1.5$ and
$\beta=0.375\pm 0.005$ in one spatial dimension. Krug \cite{K94_PRL.72.2907} 
studied the height-height correlation function and structure factor in one
dimension by taking the solid-on-solid (SOS) model with DT
\cite{sarma-tamborenea_91} relaxation rule  where overhangs are not present. His
obtained growth exponent is $\beta=0.37\pm0.01$, satisfying the scaling relation
of equation \ref{Mullins}. The step size distribution of his study yields a
non zero skewness in one substrate dimension. This implies that the 
distribution of height fluctuations is not Gaussian unlike that of equation \ref{Mullins} and consequently 
suggesting that the continuum model for the MBE growth process must be a nonlinear dynamical 
equation. 

Sun, Guo and Grant (SGG) \cite{PhysRevA.40.6763} proposed a nonlinear
equation for surface growth where  total volume under the
interface is conserved, including noise conservation. They obtained a
distinct set of exponents given by $\chi=\frac{2-d}{3}$ and $z=\frac{10+d}{3}$
that agree with the conserved restricted
solid on solid (CRSOS) model for $d=1$. 

Considering a geometrical
interpretation,
Villain, Lai and Das Sarma (VLDS) \cite{Villain_j_Phys_91,VLDS_91} proposed a nonlinear equation 
with non-conserved noise for the MBE process. This dynamical equation is known as the VLDS equation, written as 
\beq
\frac{\partial h} {\partial t}=-\nu_0
\nabla^{4} h+\frac{\lambda_0}{2}\nabla^2(\nabla h)^{2}+ \eta,
\label{VLDS}
\eeq
where $h(\vx,t)$ is the height of the fluctuating surface at position $\vx$ and
time $t$, $\nu_0$ is the diffusion constant and $\eta(\vx,t)$ is a Gaussian white 
noise with zero mean and covariance
\beq
\langle \eta(\vx,t) \eta(\vx',t')\rangle=2 D_0 \delta^d(\vx-\vx')\delta(t-t').
\eeq
where $D_0$ is a constant. 

Family and Vicsek \cite{Family_Physica} proposed the dynamic scaling for the surface width (standard deviation) as
\beq
w(L,t) \sim L^{\chi} F\left(\frac{t}{L^z}\right).
\eeq
For large $\tau$, $F(\tau)=Const.$ and for small $\tau$,  $F(\tau)\sim \tau^{\beta}$. 
The ratio $\beta=\frac{\chi}{z}$ is called the growth exponent.

Through one-loop dynamic RG calculations \cite{VLDS_91}, Lai and Das Sarma as
well as Tang and Nattermann \cite{TN91_PRL.66.2899} independently obtained the
roughness, dynamic and growth  exponents as $\chi=\frac{4-d}{3}$,
$z=\frac{8+d}{3}$ and $\beta=\frac{4-d}{8+d}$. Janssen argued in the Ref.\cite{Jan97_PRL.78.1082} 
that KPZ like Galilean transformation in SGG and VLDS equations are mathematically ill defined and 
thereby the exponents following from an RG calculations are to be corrected. Accounting for two-loop 
corrections in an RG calculation, he obtained small corrections to the exponents in each dimensions. 
Subsequently, numerical works \cite{YoKK97_PRE.56.4085,YoLK98_PRE.58.5150} were devoted to 
capture corrections to the exponents. These numerical corrections turned out to be larger than Janssen's
corrections. However, no conclusions could be reached regarding the role of vertex corrections in their 
numerical simulations. Katzav \cite{Kat02_PRE.65.032103} investigated equation \ref{VLDS} via a 
self-consistent-expansion method which has close relation to mode-coupling approach and obtained 
the same set of exponents as the one-loop dynamic RG results. It may however be noted that Katzav's 
self-consistent integral equations were equivalent to a renormalized perturbation theory at 
one-loop order without vertex renormalization.

There exist two types of investigations in the literature. One of
them is finding the scaling exponents through different growth models and
comparing them with the calculated exponents corresponding to the continuum
equations. The other type of work is the derivation of a continuum equation from
the discrete model using the principle of symmetry or re-parametrization of
invariance \cite{RevModPhys.68.963} (alternatively using the master equation
\cite{vvedensky_93,PhysRevE.54.3933,PhysRevE.54.5935,PhysRevE.57.4480}).
Wilby \emph{et al.} \cite{WVZ92_PRB.46.12896} studied the SOS model by using
Monte-Carlo-Simulation and estimated the growth exponents $\beta=0.333\pm0.010$
in $d=1$ dimension which agrees with the dynamic RG result. A conserved growth
model with RSOS condition, in $1+1$ dimensions, has been studied by Kim and
Kim in the Refs. \cite{Kim94_JPA.27.L553,Kim97_PRE.55.3977} and their estimated
exponents (roughness and growth exponents $\chi=0.95\pm0.04$ and
$\beta=0.32\pm0.01$ respectively) agree with the calculated exponents from
dynamic RG.  

There exist many numerical studies to obtain the scaling exponents which in turn determine the 
universality classes. A detailed study of the height fluctuations involves the determination of its probability 
distribution function that determines the universality class of the growth process. 
However, obtaining the full probability distribution function is an analytically impossible task and 
alternatively a few lower order moments yielding the skewness and kurtosis could be studied. The $n$th moment of the 
height fluctuations is defined as
\beq
W_n=\langle[h(\vx,t)-\bar{h}(t)]^n\rangle,
\label{nthmoment}
\eeq
 and the moments follow the power laws at stationary state as, 
 $W_n\sim L^{n\chi}$, where $L$ is the substrate size. Skewness is defined as
\beq
S=\frac{W_3}{(W_2)^{3/2}}.
\label{skew}
\eeq
It may be noted that numerical attempts to obtain the skewness for the VLDS type growth 
is rather rare. For example, Das Sarma \emph{et al.} \cite{Das-Sarma_PRE53_1996} considered a discrete 
LDS model and obtained $S=-0.1\pm 0.15$ in the steady state. 

However, there appears to be no attempts to obtain the value of skewness via analytical means directly 
from the VLDS equation. Since skewness is an important property of the probability distribution function,
it is indispensable to calculate its value for the height fluctuations governed by the VLDS dynamics 
so that this fluctuations can be distinguished from those governed by other dynamical processes. 

In this paper we calculate the second and third moments of the height fluctuations in the stationary 
state directly from the VLDS equation starting from flat initial condition employing a diagrammatic 
methods and a renormalization scheme without rescaling. This allows for the calculation of the 
skewness given by equation \ref{skew}. 

The rest of the paper is organized in the following way. Section II presents  a renormalization group scheme without
rescaling. Section III presents the calculation of the second moment $W_2$. In Section-IV, we calculate the
third moment $W_3$ and thereby the value of skewness is obtained. Section-V calculates the skewness and Section-VI gives 
a Discussion and Conclusion.

\section{Renormalization Scheme Without Rescaling}
In this section, we employ an RG scheme without rescaling that was previously used by Yakhot and Orszag 
\cite{yakhot_j_s_comput_1_3_86, yakhot_prl_57_1772_86} 
in the case of fluid turbulence and, later, it was applied to the stochastic KPZ dynamics 
\cite{Singha-Nandy14_PRE.90.062402,Singha-Nandy15_jstatmech}
We begin with the Fourier transformation of the VLDS equation (equation \ \ref{VLDS}), namely,  
\beq
(-i \omega+\nu_0 k^4) h(\vk,\omega)=\eta(\vk,\omega)-  \frac{\lambda_0}{2} k^2 
\int\!\!\int\frac{d^d q
d\Omega}{[2\pi]^{d+1}}[\vq\cdot(\vk-\vq)]h(\vq,\Omega)h(\vk-\vq,\omega-\Omega). 
\label{FTVLDS}
\eeq
We shall use equation \ref{FTVLDS} to obtain the renormalized surface diffusivity 
as well as the skewness of the height fluctuations in $(1+1)$-dimensions. 

\subsection{Scale Elimination}
In the Fourier space, we split the height and noise fields into slow and fast modes, namely, $h^<(\vk,\omega)$
and $h^>(\vk,\omega)$, and  $\eta^{<}(\vk,\omega)$ and $\eta^{>}(\vk,\omega)$, where 
$h^{>}(\mb k, \omega)$ and $\eta^{>}(\mb k, \omega)$ belong to the shell $\Lambda_0 e^{-r} \leq k \leq \Lambda_0$.
The fluctuating modes  $h^>(\mb k,\omega)$
are eliminated by means of integration and its effect is accounted for as corrections to the model parameters 
($\nu_0$, $D_0$ and $\lambda_0$). As a consequence, we obtain an equation for the slow modes $h^{<}(\mb k, \omega)$ 
\begin{eqnarray}
&& [-i \omega+\nu_0 k^4+\Sigma(\vk,\omega)] h^{<}(\vk,\omega)= \eta^{<}(\vk,\omega)
\nonumber \\ 
&& - \left(\frac{\lambda_0}{2}\right) k^2 \int\!\!\int\frac{d^d q \ d\Omega}{[2\pi]^{d+1}}[\vq
\cdot(\vk-\vq)]h^{<}(\vq,\Omega)h^{<}(\vk-\vq,\omega-\Omega), 
\label{elimination}
\end{eqnarray}
in the momentum space $0 \leq (k,p,q) \leq \Lambda_0 e^{-r}$.
Here, the self energy correction is given by  
\begin{eqnarray}
&& \Sigma(k,\omega)= 4\left(\frac{-\lambda_0}{2}\right)^2  (2D_0) k^2 \int^{>} \frac{d^d
q}{[2\pi]^d} [\vq \cdot (\vk-\vq)]  \ [\vk \cdot (\vq-\vk)] \ q^2
 \nonumber \\ 
&& \times \int^{\infty}_{-\infty} \frac{d\Omega}{[2\pi]}  \, G^{>}_0(\hat{q}) \, |G^{>}_0(\hat{k}-\hat{q})|^2
\label{efct-elimn}
\end{eqnarray}
with $G_0(\hat{k})\equiv G_0(\vk,\omega)=[-i \omega+\nu_0 k^4]^{-1}$ is a bare propagator.
Following Ref.\cite{forster_pra_16_732_77,yakhot_j_s_comput_1_3_86,yakhot_prl_57_1772_86}, 
we integrate over the internal frequency $\Omega$
and invoke the transformation $\vq \rightarrow (\vq+\vk/2)$.  
Performing the internal momentum integration in the shell $\Lambda_0 e^{-r} \leq q \leq \Lambda_0$, 
a correction to the surface diffusivity is obtained as
\beq
\Delta \nu = \frac{\lambda^2_0 D_0}{\nu^2_0 } \frac{S_d}{[2\pi]^d}
\left(\frac{6-d}{4d}\right)
\left[\frac{e^{(4-d)r}-1}{(4-d)\Lambda^{4-d}_0}\right]
\label{cortion_nu}
\eeq
in the large-scale ($q>>k$) long-time limits ($\omega \rightarrow 0$). Here $S_d=\frac{2\pi^{d/2}}{\Gamma (d/2)}$ 
is the area of a sphere of unit radius embedded in a $d$-dimensional space and $\Sigma(\vk,0)=k^4 \Delta \nu$.
Consequently, we
obtain an effective diffusivity as
\beq
\nu^{<}(r)=\nu_0 \left[1+ \frac{\lambda^2_0 D_0}{\nu^2_0 } \frac{S_d}{[2\pi]^d}
\left(\frac{6-d}{4d}\right) \left(\frac{e^{(4-d)r}-1}{(4-d)\Lambda^{4-d}_0} \right) \right ].
\label{eq-dnu}
\eeq

The two point height-height correlation function corresponding to one-loop
perturbative expansion can be written as 
\begin{equation}
\langle h^{<}(\vk,\omega) h^{<}(\vk',\omega') \rangle= 2[D_0+\Delta D]
|G(\vk,\omega)|^2  
[2\pi]^{d+1} \delta^d(\vk+\vk') \delta(\omega+\omega')
\label{height_corlan}
\end{equation}
where 
\beq
\Delta D=  \left(\frac{-\lambda_0}{2}\right)^2 (2D_0)^2 k^4 \int^{>}
\frac{d^d q}{[2\pi]^d} \, [\vq \cdot (\vk-\vq)]^2 \int^{\infty}_{-\infty} \frac{d
\Omega}{[2\pi]} \, |G^{>}_0(\hat{q})|^2 \, |G^{>}_0(\hat{k}-\hat{q})|^2
\label{correlation}
\eeq
which comes from the amputated part of the loop diagram shown in figure \  1(b).
The frequency and momentum integrations in equation \ref{correlation} are 
carried out yielding
\beq
\Delta D= k^4 \left(\frac{K_d \lambda^2_0 
D^2_0}{4\nu^3_0}\right) \left[\frac{\Lambda_0^{d-8}-\Lambda^{d-8}(r)}{d-8}
\right]
\eeq
Since $\Delta D$ goes like $k^4$, this correction to $D_0$ is irrelevant in the large-scale and long-time limits. 
Hence \beq
D^{<}(r)=D_0 .
\label{eq-D0}
\eeq

According to Lai and Das Sarma \cite{VLDS_91}, the
vertex ($\lambda_0$) does not get any corrections from the RG transformation at one-loop order, 
consequently \beq
\lambda^{<}(r)=\lambda_0 .
\label{eq-lambda0}
\eeq

\subsection{Flow equations and Fixed point}
We obtained the flow equations as differential equations with respect to variation of the parameter $r$ from
equations \ref{eq-dnu}, \ref{eq-D0} and \ref{eq-lambda0} giving
\begin{eqnarray}
\frac{d\nu}{dr}&=&K_d \frac{\lambda_0^2
D_0\Lambda^{d-4}(r)}{\nu^2(r)} \left(\frac{6-d}{4d}\right) \nonumber \\
\frac{dD}{dr}&=& 0 \nonumber  \\
\frac{d\lambda}{dr}&=& 0 \nonumber \\
\label{flow_equation}
\end{eqnarray}
where $\Lambda(r)=\Lambda_0 e^{-r}$.
We define an effective coupling parameter 
\beq
g(r)=K_d \frac{\lambda^2_0 D(r)}{\nu^3(r) \Lambda^{4-d}(r)}.
\label{g(r)} 
\eeq
Using  equations \ \ref{flow_equation} and  \ref{g(r)}, we obtain the flow equation for the coupling constant 
\beq
\frac{dg(r)}{dr}=a\ g(r)-b\ g^2(r)
\label{difg(r)}
\eeq
where $a=(4-d)$ and $b=\frac{3(6-d)}{4d}.$ Integrating equation \ref{difg(r)} over $r$, yields 
\beq
g(r)=\frac{g_0 e^{a r}}{1+\frac{b}{a} g_0 \ (e^{a r}-1)},
\label{solg(r)}
\eeq
where $g(0)=g_0=\frac{K_d \lambda^2_0 D_0}{\nu_0^3 \Lambda_0^{4-d}}.$ In the asymptotic limit of 
$r$ ($r \rightarrow \infty$),  equation \ref{solg(r)} approaches to $r$ independent quantity
\beq
g*=\frac{4d(4-d)}{3(6-d)}.
\label{FP}
\eeq
The fixed point, expressed by equation \ref{FP}, is stable for $d<4$, thereby the upper critical
dimension of the VLDS equation is $d_c=4$ which agrees with Refs. \cite{VLDS_91,TN91_PRL.66.2899}.

\subsection{Renormalized Surface Diffusivity}
Integrating the first flow equation in equation \ref{flow_equation} using 
equation \ref{difg(r)}, we obtain 
\beq
\nu(r)=\nu_0 \left[1+\frac{b}{a} g_0
\left(e^{ar}-1\right)\right]^{1/3}.
\label{nur}
\eeq
In the asymptotic limit of large $r$, equation \ref{nur} takes the form 
\beq
\nu(r)=\nu_0 \left(\frac{b\ g_0}{a} e^{a r}\right)^{1/3},
\eeq
and, in $(1+1)$ dimensions, the above equation becomes 
\beq
\nu(r)=\nu_0 \left(\frac{5\ g_0}{4}\right)^{1/3} e^{r}.
\label{nurd1}
\eeq
Thus the corresponding renormalized surface diffusivity reads
\begin{eqnarray}
\nu(k)=&& \nu_0 \left(\frac{5
\lambda^2_0 D_0}{4\pi \nu^3_0} \right)^{1/3} k^{-1},
\label{nukd1}
\end{eqnarray}
in the large-scale long-time limits. The renormalized response function $G(\vk,\omega)$ and the dynamic exponent $z$
is related through the expression
\beq
G^{-1}(\vk,\omega)=[-i \omega+ \nu(k) k^4] \  \propto \ 
k^z \, \xi\left(\frac{\omega}{k^z}\right).
\label{resdynarela}
\eeq
The above equation indicates that $\nu(k) k^4\sim k^z$ implying dynamic
exponent  $z=3$ in $d=1$ dimension. The roughness exponent, $\chi=1$, can be obtained 
from the scaling relation $\chi+z=4$.  

\section{The Second Moment}
Using the definition of $n$th moment of the height fluctuations expressed in equation \ref{nthmoment}, the
second and third moments are obtained as 
\beq
W_2=\langle h^2(\vx,t)\rangle - \langle h(\vx,t) \rangle^2 .
\label{eq-W2}
\eeq
and 
\beq 
W_3=\langle
h^3(\vx,t)\rangle-3 \langle h^2(\vx,t) \rangle \langle h(\vx,t)\rangle
+ 2 \langle h(\vx,t)\rangle ^3,
\label{eq-W3}
\eeq
respectively.
Here $h(\vx,t)$ is the height fluctuations satisfying $\langle h(\vx,t)\rangle=0$, and consequently  
equation \ref{eq-W2} and equation \ \ref{eq-W3} become
\beq
W_2=\langle h^2(\vx,t)\rangle 
\eeq
and 
\beq 
W_3=\langle h^3(\vx,t)\rangle,
\eeq
\subsection{Calculation of $W^{(1)}_2$}
The perturbative expansion at one-loop order of the second moment yields
\beq 
W_2=\langle [h(\vx,t)]^2 \rangle =W_2^{(1)}+W_2^{(2)}+W_2^{(3)}, 
\label{petur-W2}
\eeq
where $W^{(1)}_2$ and $W^{(2)}_2$ are non-zero contributions corresponding to 
$O[\lambda^{0}_0]$ and $O[\lambda^2_0]$ of the perturbation series.
In Fourier space, the expression for $W^{(1)}_2$ is written as 
\begin{equation}
 W^{(1)}_2= \int \frac{d^{d}k d \omega}{[2\pi]^{d+1}}  
\int \frac{d^{d}k' d \omega'}{[2\pi]^{d+1}} Q^{(0)}(\mb k,\omega;\mb k',\omega') e^{i(\vk+\vk') \cdot \vx} 
e^{-i(\omega+\omega')t}.
\end{equation}
The diagrammatic representation of $W^{(1)}_2$ is shown in figure 1(a). We incorporate two point 
correlation (zeroth order), namely,  
$Q^{(0)}(\mb k,\omega;\mb k',\omega') \equiv \langle h(\mb k,\omega) h(\mb k',\omega')\rangle=2 D_0 [2\pi]^{d+1}  |G(\mb k, \omega)|^2 \delta^d(\mb k+\mb k') \delta(\omega+\omega')$
and carry out integration over $k'$ yielding
\begin{equation}
W^{(1)}_2= 2D_0 \int \frac{d^{d}k}{[2\pi]^{d}} \int \frac{d \omega}{[2\pi]}  |G(\mb k, \omega)|^2.
\label{W12-bef-fre-mom}
\end{equation}
We use the functional form of $\nu(k)$ from equation \ref{nukd1}. 
Consequently performing the frequency integration and carrying out the 
momentum integration in $(1+1)$ dimensions in equation \ref{W12-bef-fre-mom}, we obtain 
\begin{equation}
W^{(1)}_2= \frac{1}{2 \pi^{2/3}} \left(\frac{4 D^2_0}{5 \lambda^2_0}\right)^{1/3} \frac{1}{\mu^2}
\label{W12-final}
\end{equation}
where $\mu$ is the lower limit of the momentum integration. 

\subsection{Calculation of $W^{(2)}_2$}
The expression for $W^{(2)}_2$ is expressed in momentum and frequency spaces as
\begin{equation}
W^{(2)}_2=\int \frac{d^dk d \omega}{[2\pi]^{d+1}} \int
\frac{d^dk' d\omega'}{[2\pi]^{d+1}} \, Q^{(2)}(\mb k,\omega;\mb k',\omega') \,  e^{i(\vk+\vk') \cdot \vx}
e^{-i(\omega+\omega')t}.
\label{h^2}
\end{equation}
where $Q^{(2)}(\mb k,\omega;\mb k',\omega')= [2\pi]^{d+1} L_2(\mb k,\omega) |G(\mb k, \omega)|^2 \delta^d(\mb k+\mb k') \delta(\omega+\omega')$. 
Feynman diagram for $W^{(2)}_2$ is given in figure \ 1(b). 
One-loop contribution to the second moment is given by 
\beq
W^{(2)}_2 = \int \frac{d^dk d \omega}{[2\pi]^{d+1}} G(\vk, \omega) L_2(\vk,\omega)
G(-\vk, -\omega)
\label{NW2}
\eeq
where $L_2$ [$L_2=k^4 l_2(\hat{k})$] corresponds to the amputated loop of figure\ 1(b) 
expressed in equation\ \ref{correlation}. We carry out the integrations over the internal 
frequencies and internal momenta 
in the shell $\Lambda_0 e^{-r} \leq k \leq \Lambda_0$ and obtain 
\beq
l^{<}_2(r)=\left(\frac{ \lambda^2_0
D^2_0}{2 \pi \nu^3_0}\right) \left[\frac{\Lambda_0^{-7}-\Lambda^{-7}(r)}{-7}
\right]
\label{L2genD}
\eeq
leading to the differential equation
\beq
\frac{dl_2}{dr}=\frac{\lambda_0^2  D_0^2}{2 \pi \nu^3(r)
\Lambda^{7}(r)}.
\label{L2(r)}
\eeq

\begin{figure}[h]
\vskip-0.1in
\hskip0.3in \includegraphics[scale=0.4]{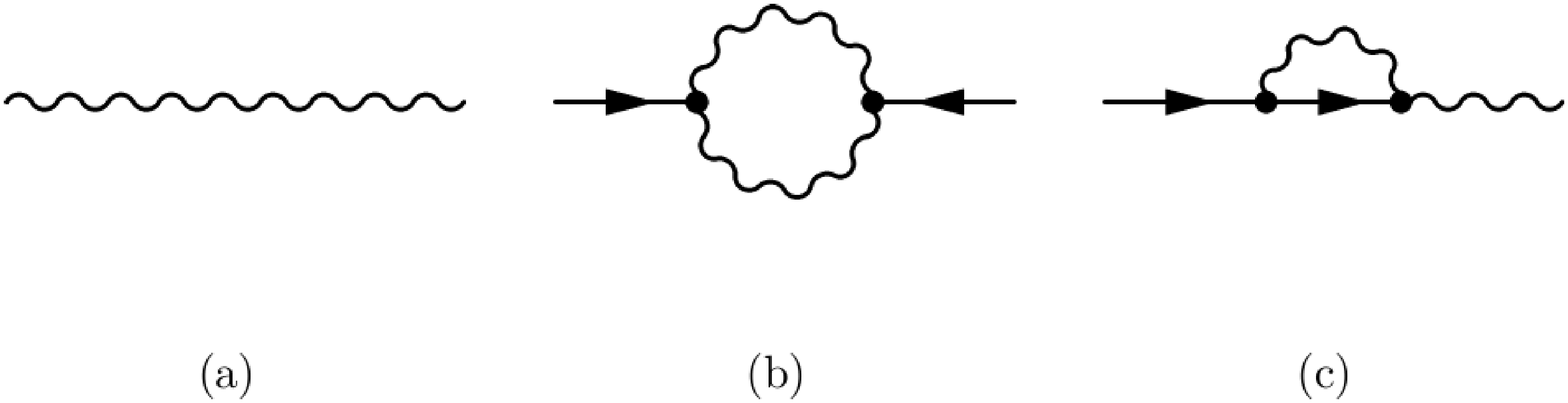}
\vskip-0.1in
\caption{Feynman diagrams : (a) $W^{(1)}_2$, (b) $W^{(2)}_2$ and (c) $W^{(3)}_2$. 
Responses are indicated by solid lines with arrow and correlation by a wiggly line.}
\label{fig2}
\end{figure}
The function $\nu(r)$ being known from
equation \ref{nurd1}, the above differential equation is solved to obtain

\beq
l_2(r)=\left(\frac{D_0}{10 \Lambda^4_0}\right) e^{4r}
\label{L_2}
\eeq 
in the asymptotic limit of large $r$. Identifying $\Lambda_0 e^{-r}=k$, we get  
\beq
l_2(k,0)=\frac{1}{10}D_0 \ k^{-4}.
\eeq

The vertex of figure\ 1(b) contains $k^4$ 
and  consequently the loop contribution (without the external legs) becomes we define $L_2(k,0)=k^4 l_2(k,0)$ 
\beq
L_2= \frac{1}{10} D_0.
\label{VertAL}
\eeq
which is scale independent.
Now substituting equation \ref{VertAL} in equation \ref{NW2}, we get 
\beq
W^{(2)}_2=\int \frac{d^d k}{(2\pi)^d} \int \frac{d\
\omega}{[2 \pi]} |G(\vk, \omega)|^2  L_2
\label{W2L2}
\eeq
Performing the frequency integration in equation\ \ref{W2L2}, we obtain
\beq
W^{(2)}_2=\frac{1}{20 \pi^{2/3}}  \left(\frac{4 D^2_0}{5\lambda^2_0}\right)^{1/3}
\int^{\infty}_{\mu} dk k^{-3}. 
\eeq
We carry out the momentum integration in the above equation leading to
\beq
W^{(2)}_2=\frac{1}{40 \pi^{2/3}} \left(\frac{4 D^2_0}{5\lambda^2_0}\right)^{1/3}
\frac{1}{\mu^2}.
\label{W22-final}
\eeq
Feynman diagram corresponding to $W^{(3)}_2$ is shown in figure 1(a).
The contribution to $W^{(3)}_2$ vanishes in the large-scale limit. Adding the contributions $W^{(1)}_2$ and 
$W^{(2)}_2$ from equations \ref{W12-final} and \ref{W22-final},  we obtain $W_2$ as
\begin{equation}
W_2= W^{(1)}_2+ W^{(2)}_2 = \frac{21}{40 \pi^{2/3}} \left(\frac{4 D^2_0}{5\lambda^2_0}\right)^{1/3} \frac{1}{\mu^2}
\label{W2-final}
\end{equation}

\section{The Third Moment and Skewness}
The third moment of height fluctuations is expressed as 
\beq
W_3 =\sum^{8}_{i=1} W^{(i)}_3
\label{petur-W3}
\eeq
where $W^{(1)}_3$ and $W^{(2)}_3$, corresponding to Feynman diagrams Figs.\ 2(a) and 2(b) respectively, give non-zero 
contributions. The terms $W^{(3)}_3$ and $W^{(4)}_3$, corresponding to figures 3(a) and 3(b), respectively, 
give equal magnitudes of logarithmic corrections with opposite signs and thus they cancel each other. 
The rest of the terms in equation \ref{petur-W3}, namely, $W^{(5)}_3$, $W^{(6)}_3$, $W^{(7)}_3$ and $W^{(8)}_3$, 
are depicted in Figs. \ 4(a), 4(b), 4(c) and 4(d), respectively, which give zero contributions individually in the 
large-scale long-time limits.


\subsection{Calculation of $W^{(1)}_3$}
In this subsection, we calculate Feynman diagram given in figure \ 2(a).
The integral expression for $W^{(1)}_3$ in the Fourier space is given by
\begin{eqnarray}
\lefteqn{
W^{(1)}_3=
\int\frac{d^dk_1\,d\omega_1}{[2\pi]^{d+1}}
\int\frac{d^dk_2\,d\omega_2}{[2\pi]^{d+1}}
\int\frac{d^dk_3\,d\omega_3}{[2\pi]^{d+1}} \nonumber}\\
&& \langle h(\vk_1,\omega_1)\,h(\vk_2,\omega_2)\,h(\vk_3,\omega_3)\rangle
\,\,e^{i(\vk_1+\vk_2+\vk_3) \cdot \vx}
\,e^{-i(\omega_1+\omega_2+\omega_3)t},
\label{W13}
\end{eqnarray}
The third order  height correlation 
appearing in equation \ref{W13} is expressed as 
\begin{eqnarray}
\langle h(\vk_1,\omega_1)\,h(\vk_2,\omega_2)\,h(\vk_3,\omega_3)\rangle=  && G(\hat{k}_1) \, G(\hat{k}_2) \, G(\hat{k}_3)
\, L^{(1)}_3(\hat{k}_1;\hat{k}_2;\hat{k}_3) \nonumber \\
&& \times [ (2\pi)^{d+1} \,\delta^{d+1}(\hat{k}_1+\hat{k}_2+\hat{k}_3) ]
\label{3height-corre-W13}
\end{eqnarray}
where $L^{(1)}_3(\hat{k}_1;\hat{k}_2;\hat{k}_3)=\mb k^2_1 \, \mb k^2_2 \, \mb k^2_3 \, l_3(\hat{k}_1;\hat{k}_2;\hat{k}_3)$ 
represents the renormalized amputated part of the loop diagram [figure \ 2(a)]. 
Substituting equation \ref{3height-corre-W13}, in equation \ref{W13}, we obtain
\begin{equation}
W^{(1)}_3=\int \frac{d^{d+1}\hat{k}_1}{[2\pi]^{d+1}} \int
\frac{d^{d+1}\hat{k}_2}{[2\pi]^{d+1}} \, G(\hat{k}_1) \, G(\hat{k}_2)
\, L^{(1)}_3(\hat{k}_1;\hat{k}_2) \,
G(-\hat{k}_1-\hat{k}_2)
\label{W32}
\end{equation}
The bare value of $l^{(1,0)}_3(\hat{k}_1;\hat{k}_2)$ is given by 
\begin{eqnarray}
l^{(1,0)}_{3}(k_1, \omega_1;k_2, \omega_2)= 8 && \left(\frac{-\lambda_0}{2} \right)^3 \int
\frac{d^dq \ d\omega}{[2\pi]^{d+1}} [(\vq-\vk_1)\cdot(\vk_2+\vk_1-\vq)]  
[\vq \cdot (\vk_1-\vq)]  
\nonumber \\
&& \times [-\vq \cdot(\vq-\vk_2-\vk_1)]  Q_0(\hat{\vq}) Q_0(\hat{\vk}_1-\hat{\vq})
Q_0(\hat{\vk}_1+\hat{\vk}_2-\hat{\vq})]. \nonumber \\
\label{vlds_loop}
\end{eqnarray}

\begin{figure}[ht]
\vskip-0.1in
\hskip0.9 in \includegraphics[scale=.54]{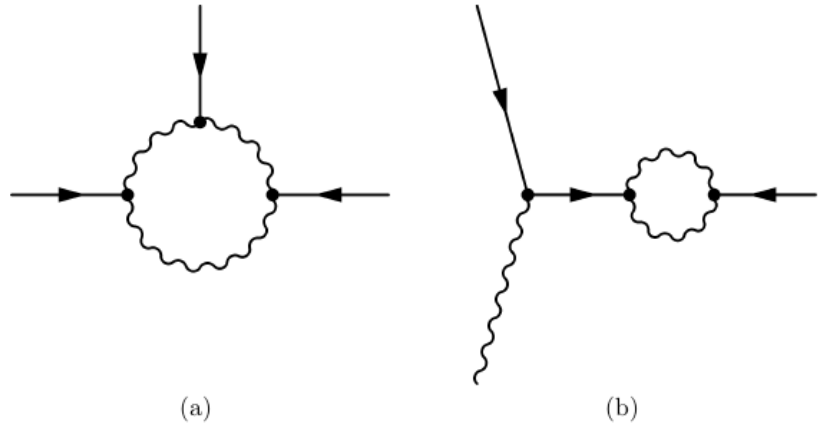}
\vskip-0.3in
\caption{Feynman diagrams: (a) $W^{(1)}_3$ and (b) $W^{(2)}_3$. These diagrams contribute to $W_3$.}
\label{fig2}
\end{figure}
where $Q_0(\vq)=2D_0 \, [2 \pi]^{d+1} \, G_0(\hat{q}) \, G_0(\hat{q}') \, \delta^{d+1}(\hat{q}+\hat{q}')$ are
the correlations. Considering the external momenta and frequencies to be much smaller
in magnitude than internal momenta and frequencies, we carry out frequency convolution
and momentum integrations yielding 
\beq
l^{(1)<}_3(r)=\frac{3}{2}  \left[\frac{\lambda^3_0 D^3_0 K_d}{\nu^5_0
\Lambda^{14-d}_0}\right] \left[\frac{e^{(14-d)r}-1}{(14-d)}\right].
\eeq
The iterative nature of the shell elimination scheme yields  
\beq
\frac{dl^{(1)}_3}{dr}=\frac{3}{2}\left[\frac{\lambda^3_0 D^3_0
K_d}{\nu^5(r) \Lambda^{14-d}(r)}\right]
\label{dif-L3}
\eeq
Solving this equation in the asymptotic limit of large $r$ in $d=1$ dimension, we obtain
\beq
l^{(1)}_3(r)=\frac{3}{20} \lambda_0 D^2_0 \left(\frac{4\pi}{5\lambda^2_0
D_0}\right)^{2/3} \frac{1}{\Lambda^{8}(r)}.
\label{L3r}
\eeq
$\Lambda_0 e^{-r}$ is identified as the external momenta, yielding  
\beq
l^{(1)}_3(k_1,0;k_2,0)= \frac{3}{20} \lambda_0 D^2_0 \left(\frac{4\pi}{5\lambda^2_0
D_0}\right)^{2/3} k_1^{-4} k_2^{-4}.
\label{l3-momentum-devide}
\eeq
So, $L^{(1)}_3$ becomes 
\begin{equation}
L^{(1)}_3(\vk_1,\vk_2)= \mb k^2_1 \, \mb k^2_2 \, \mb k^2_3 \, l^{(1)}_3
\end{equation}
Now to find the wave vector and frequency dependence in equation \ref{l3-momentum-devide} a form of scaling function is
introduced by replacing $k^{-2}_{i}$ by 
\beq 
k_i^{-2}f_d\left(\frac{\omega_i}{k^z_i}\right) = k_i^6 \nu^2(k_i)
|G(\vk_i,\omega_i)|^2 
\label{dynamical-scaling}
\eeq
where $i=1,2$ and $f(0)=1$. This expression is the same as $k^{-2}_{i}$ for $\omega_i=0$.
Incorporating equation \ref{dynamical-scaling} in equation \ref{l3-momentum-devide}, we obtain 
\begin{eqnarray}
L^{(1)}_3(\hat{k}_1,\hat{k}_2)=\frac{3}{20} A \ k_1^4   k_2^4|\vk_1+\vk_2|^2
|G(\hat{k}_1)|^2 |G(\hat{k}_2)|^2 
\label{L'-final}
\end{eqnarray}
with $A= \lambda_0 D^2_0 \left(\frac{5\lambda^2_0 D_0}{4\pi}\right)^{2/3}$.
We substitute equation \ref{L'-final} in equation \ref{W32} yielding  
\begin{eqnarray}
W^{(1)}_3= \frac{3}{20} A \int \frac{d^{2}\hat{k}_1}{[2\pi]^{2}}
\int \frac{d^{2}\hat{k}_2}{[2\pi]^{2}} && \, k_1^4 \, k^4_2 \, |\vk_1+\vk_2|^2
   \ |G(\hat{k}_1)|^2 \, |G(\hat{k}_2)|^2  \nonumber \\
 && \times [G(\hat{k}_1) G(\hat{k}_2) \ G(-\hat{k}_1-\hat{k}_2)] 
\label{W3-preinte}
\end{eqnarray}
Carrying out the frequency integrations over $\omega_1$ and $\omega_2$, we obtain 
\beq
W^{(1)}_3=\left(\frac{3}{20}\right) \frac{A_1}{[2\pi]^2}
\int^{\infty}_{-\infty} dk_1 \!\int^{\infty}_{-\infty}
dk_2 \ F_1(k_1,k_2) . 
\label{W3f}
\eeq
where 
\begin{equation}
\hskip-.85in
F_1(k_1,k_2)=\frac{
3\sigma^2(k_1)+(\sigma(k_2)+\sigma(|k_1+k_2|)(
3\sigma(k_2)+\sigma(|k_1+k_2|))
+2\sigma(k_1)(7\sigma(k_2)+2\sigma(|k _1+k_2|))} {
16 \sigma^2(k_1)\sigma^2(k_2)(\sigma(k_1)+\sigma(k_2)+\sigma(|k_1+k_2|))^3} \nonumber \\
\end{equation}
with $\sigma(k_{i})= k_i^3$ and $A_1=\frac{A \ (4\pi)^{5/3}}{\left(5\lambda^2_0
D_0\right)^{5/3}}=\left(\frac{4 \pi D_0}{5\lambda_0}\right) $. 
 Now the integrations become 
$$ I_1(\mu)= \int^{\infty}_{\mu} dk_1 \int^{\infty}_{\mu} dk_2 F_1(k_1,k_2)$$
and
$$ J_1(\mu)=\int^{\infty}_{\mu} dk_1  \int^{\infty}_{\mu} dk_1  F_1(-k_1,k_2).$$
Since these integrals are infrared divergent, we write them as 
\begin{eqnarray}
I_1(\mu)=I^{(0)}_1 \mu^{-3} \nonumber \\
J_1(\mu)=J^{(0)}_1 \mu^{-3}.
\label{int-I0J0}
\end{eqnarray}
where $I_1^{(0)}$ and $J^{(0)}_1$ are dimensionless numerical constants. 
Carrying out numerical integrations, we obtain the values 
for the constants as
\begin{eqnarray}
I_1^{(0)}=\lim_{\mu \rightarrow 0^+} [\mu^{3} I_1(\mu)]= 0.0148673 \nonumber \\
J_1^{(0)}=\lim_{\mu \rightarrow 0^+} [\mu^{3} J_1(\mu)]=  0.0039435 ,
\label{int-I0J0-value}
\end{eqnarray}

Substituting equations \ref{int-I0J0} and \ref{int-I0J0-value} in equation \ref{W3f}, we obtain

\begin{eqnarray}
W^{(1)}_3=&&\frac{3}{100\pi} \left(\frac{D}{\lambda_0}\right)
(2I^{(0)}_1+2J^{(0)}_1) \frac{1}{\mu^{3}} \nonumber \\
&& =\frac{3}{50\pi} \left(\frac{D}{\lambda_0}\right) (0.018811)
\frac{1}{\mu^{3}}.
\label{W31D1st-dia}
\end{eqnarray}


\subsection{Calculation of $W^{(2)}_3$}
Feynman diagram for $W^{(2)}_3$  is shown in figure 2(b). The integral expression for 
$W^{(2)}_3$ is written as  

\begin{eqnarray}
& W^{(2)}_3 & =\int \frac{d^{d+1} \hat{k}}{[2\pi]^{d+1}} \int \frac{d^{d+1}\hat{k}'}{[2\pi]^{d+1}} \int \frac{d^{d+1}\hat{k}''}{[2\pi]^{d+1}}
\int \frac{d^{d+1}\hat{q}}{[2\pi]^{d+1}} \int \frac{d^{d+1}\hat{p}}{[2\pi]^{d+1}} \int \frac{d^{d+1}\hat{Q}}{[2\pi]^{d+1}} \nonumber \\
&& [k'^2 G(\hat{k'}) k''^2 G(\hat{k''}) Q^2 G(\hat{Q})] [\mb q \cdot (\mb k'-\mb q)] [\mb Q \cdot (\mb k''-\mb Q)]
[\mb p \cdot (\mb Q-\mb p)] \nonumber \\
&& \times [\langle h(\hat{k}) h(\hat{q}) h(\hat{k'}-\hat{q}) h(\hat{k''}-\hat{q}) h(\hat{p})  h(\hat{Q}-\hat{p}) \rangle]
\label{W23-initial}
\end{eqnarray}
The height correlation appearing in equation \ref{W23-initial} is expressed as 
\begin{eqnarray}
&&\langle h(\hat{k}) h(\hat{q}) h(\hat{k'}-\hat{q}) h(\hat{k''}-\hat{Q}) h(\hat{p}) h(\hat{Q}-\hat{p}) \rangle = 24 (2 D_0)^3 
[2 \pi]^{3(d+1)} [|G_0(\hat{q})|^2] |G_0(\hat{k'}-\hat{q})|^2 \nonumber \\
&& \times |G(\hat{k})|^2 [G(\hat{k'}) k'^2] [G(-\hat{k}-\hat{k'}) |\mb k+\mb k'|^2] [\mb k'^2 G(-\hat{k'})] \delta^{d+1}(\hat{k}+\hat{k'}+\hat{k''}) 
[\delta^{d+1}(\hat{p}+\hat{q})] \nonumber \\
&& \times [\delta^{d+1}(\hat{Q}+\hat{k'})] 
\end{eqnarray}
Substituting the above form in equation \ref{W23-initial}, we obtain
\begin{eqnarray}
& W^{(2)}_3 & = 12 (\lambda_0 D_0 ) \int \frac{d^{d+1}\hat{k}}{[2\pi]^{d+1}} \int \frac{d^{d+1}\hat{k}'}{[2\pi]^{d+1}} 
|\mb k+ \mb k'|^2 [\mb k' \cdot (-\mb k)] |G(\hat{k'})|^2 |G(\hat{k})|^2 G(\hat{k}-\hat{k'})  \nonumber \\ 
&&  L^{(2)}_3(\hat{k'})
\label{W23-L2-intro}
\end{eqnarray}
where $L^{(2)}_3(\hat{k'})=k'^4 l^{(2)}_3(\hat{k'})$ and
\begin{equation}
l^{(2)}_3(\hat{k'})= 2 (\lambda_0 D_0)^2 \int \frac{d^{d+1}q}{[2\pi]^{d+1}} [\mb q \cdot (\mb q-\mb k')]^2 
[\mb q \cdot (\mb k'-\mb q)] |G_0(\hat{k'}-\hat{q})|^2 |G_0(\hat{q})|^2
\label{L2-initial}
\end{equation}
We consider the external momenta and frequencies are much smaller in magnitude 
than internal momenta and frequencies and perform the frequency integration and momentum integration in the range 
$\Lambda_0 e^{-r} \leq q \leq \Lambda_0$ leading
\begin{equation}
l^{(2)<}_3(r) = \frac{(\lambda_0 D_0)^2 K_d}{ 2 \nu^3_0} \left[\frac{\Lambda^{d-8}_0-\Lambda^{d-8}(r)}{(d-8)}\right].
\end{equation}
We construct a differential equation for $l^{(2)}_3(r)$ by considering the iterative nature of the procedure 
\begin{equation}
\frac{d l^{(2)}_3}{dr}= \frac{\lambda^2_0 D^2_0 K_d}{2 \nu^3(r)} \Lambda^{d-8}(r) 
\label{L2-diff}
\end{equation}
Integrating over $r$, we obtain 
\begin{equation}
l^{(2)}_3(r)= \frac{D_0}{10} \Lambda^{-4}(r) 
\end{equation}
in $d=1$ dimension.
We identify $\Lambda(r)$ as  an external momentum and obtain 
\begin{equation}
 L^{(2)}_3=k'^4 l^{(2)}_3(k',0)=D_0/10
\label{L^{'(2)}_3}
 \end{equation}
which is an scale independent quantity. We substitute equation \ref{L^{'(2)}_3} in equation \ref{W23-L2-intro} leading to
\begin{equation}
W^{(2)}_3 =\frac{6}{5} (\lambda_0 D^2_0) \int \frac{d^{2}\hat{k}}{[2\pi]^{2}} \int \frac{d^{2}\hat{k}'}{[2\pi]^{2}} 
[-\mb k' \cdot \mb k] |\mb k+\mb k'|^2 |G(\hat{k}')|^2 |G(\hat{k})|^2 G(-\hat{k}-\hat{k}')
\label{W23-1D-L2-substi}
\end{equation}
Carrying out the frequency  integration, we obtain
\begin{equation}
W^{(2)}_3=-\frac{6 \pi}{25} \left(\frac{D_0}{\lambda_0}\right) \int^{\infty}_{-\infty} \frac{d k}{[2\pi]} 
\int^{\infty}_{-\infty} \frac{d k'}{[2\pi]} F_2(k,k')
\end{equation}
where 
\begin{equation}
F_2(k,k')= \frac{(k' k)(k+k')^2}{|k|^3 |k'|^3 [|k|^3+|k'|^3+|k+k'|^3]}
\end{equation}
We write 
\begin{equation}
W^{(2)}_3 =- \left(\frac{3}{50 \pi}\right) \left(\frac{D_0}{\lambda_0}\right) [2(I_2(\mu)+J_2(\mu))]
\label{W23-Imu-Jmu}
\end{equation}
where
$$ I_2(\mu)= \int^{\infty}_{\mu} dk \int^{\infty}_{\mu} dk' F_2(k,k')$$
and
$$ J_2(\mu)=\int^{\infty}_{\mu} dk  \int^{\infty}_{\mu} dk'  F_2(-k,k').$$
As before we write the integrals as
\begin{eqnarray}
I_2(\mu)=I^{(0)}_{2} \mu^{-3} \nonumber \\
J_2(\mu)=J^{(0)}_{2} \mu^{-3}.
\label{I0J0}
\end{eqnarray}
Numerical evaluations yield 
\begin{eqnarray}
I^{(0)}_2=\lim_{\mu \rightarrow 0^+} [\mu^{3} I_2(\mu)]= 0.154439 \nonumber \\
J^{(0)}_2=\lim_{\mu \rightarrow 0^+} [\mu^{3} J_2(\mu)]=-0.020075,
\label{W23I0-J0}
\end{eqnarray}
Substituting equations \ref{W23I0-J0} and \ref{I0J0} in equation \ref{W23-Imu-Jmu}, we obtain
\begin{eqnarray}
W^{(2)}_3=&&\frac{3}{50\pi} (2I^{(0)}_{2}+2J^{(0)}_{2}) \left(\frac{D}{\lambda_0}\right) \frac{1}{\mu^{3}} \nonumber \\
&& =-\frac{3}{50\pi}  (0.268728) \left(\frac{D}{\lambda_0}\right) \frac{1}{\mu^{3}}.
\label{W31D2nd-dia}
\end{eqnarray}



\subsection{Calculation of $W^{(3)}_3$}
The integral expression for $W^{(3)}_3$ can be written as
\begin{eqnarray} 
W^{(3)}_3 &=& \left(\frac{-\lambda_0}{2}\right)^3 \int \frac{d^{d+1} \hat{k}}{(2\pi)^{d+1}} 
\int \frac{d^{d+1} \hat{k}'}{(2\pi)^{d+1}}\int \frac{d^{d+1} \hat{p}}{(2\pi)^{d+1}}  
\int \frac{d^{d+1} \hat{q}}{(2\pi)^{d+1}} \int \frac{d^{d+1} \hat{q}'}{(2\pi)^{d+1}} 
\int \frac{d^{d+1} \hat{Q}}{(2\pi)^{d+1}} 
  \nonumber \\ 
  && \times [\mb Q \cdot (\mb k'- \mb Q)] [\mb q' \cdot (\mb Q- \mb q')] ]
   [\mb q \cdot (\mb q'-\mb q)][G(\hat{k}') |k'|^2  |Q^2| |q'|^2  G_0 (\hat{Q}) G_0 (\hat{q}') \nonumber \\ 
   && \times \langle h(\hat{p}) h(\hat{k}) h(\hat{k}'-\hat{Q}) h(\hat{Q}-\hat{q}') h(\hat{q}) h(\hat{q}'-\hat{q}) \rangle
 \label{W33-ini}
 \end{eqnarray}
which is depicted in figure 3(a).
 The height correlation appearing in equation \ref{W33-ini} is expressed as 
 \begin{eqnarray}
 \langle h(\hat{p}) h(\hat{k}) h(\hat{k}'-\hat{Q}) h(\hat{Q}-\hat{q}') h(\hat{q}) h(\hat{q}'-\hat{q}) \rangle &=&
 48 (2D_0)^3 [2 \pi]^{3(d+1)} |G_0(\hat{q}'+\hat{k})|^2 |G_0(\hat{p})|^2 \nonumber \\ 
 && |G_0(\hat{k})|^2 \delta^{d+1}(\hat{Q}+\hat{p}-\hat{q}') \delta^{d+1}(\hat{q}'+\hat{k}-\hat{q}) \nonumber \\ 
 && \delta^{d+1}(\hat{q}'-\hat{p}-\hat{k}')
 \label{W33-6heights-corr}
\end{eqnarray}
 Substituting equation \ref{W33-6heights-corr} in equation \ref{W33-ini}, we obtain 
\begin{eqnarray}
  W^{(3)}_3=-48 (\lambda_0 D_0)^3 \int \frac{d^{d+1} \hat{k}}{(2\pi)^{d+1}} \int \frac{d^{d+1} \hat{p}}{(2\pi)^{d+1}} 
  |\mb k+\mb p|^2 |G(\hat{p})|^2 |G(\hat{k})|^2] G(-\hat{k}-\hat{p}) L_3(\hat{k};\hat{p}) \nonumber \\
 \label{W33-ampu-short}
\end{eqnarray}
 where 
\begin{eqnarray}
 l^{(3)}_3(\hat{k};\hat{p})&=& \int \frac{d^{d+1} \hat{q}'}{(2 \pi)^{d+1}}  |\mb q'-\mb p|^2 |q'|^2 [\mb q' \cdot \mb p] 
 [(\mb q'-\mb p) \cdot (\mb k+\mb q')] [(\mb q'+\mb k) \cdot \mb k] G_0(\hat{q}') \nonumber \\
&&  G_0(\hat{q}'-\hat{p})  |G_0(\hat{q}'+\hat{k})|^2
 \label{L33-bef-renorma}
\end{eqnarray}
Assuming the internal wave vector $q'$ is much greater than external wave vectors $p$ and $k$, we get
\begin{eqnarray}
 L^{(3)}_3(\mb k,0;\mb p,0)= -\int \frac{d^{d+1} q'}{[2 \pi]^{d+1}} |q'|^6 [\mb q' \cdot \mb p] [\mb q' \cdot \mb k] |G_0(\hat{q}')|^2
 G^2_0(\hat{q}').
 \label{L33-bare}
\end{eqnarray}
 Carrying out the frequency and momentum ($\Lambda_0 e^{-r}\leq q' \leq \Lambda_0$) integration in $d=1$ dimension, 
 we follow the same procedure as in Sec. 4.1 and obtain 
\beq
L^{(3)}_3(p,0;k,0)=-\frac{p \, k}{10 \lambda^2_0 D_0} \ln(\Lambda_0/k).
\label{L33-final-form}
\eeq
We substitute equation \ref{L33-final-form}  in equation \ref{W33-ampu-short} and obtain 
 \begin{equation}
 W^{(3)}_3 =\frac{24}{5} \left(\lambda_0 D^2_0\right) \, \, I_3
 \label{W33-final-form}
 \end{equation}
 where 
 \beq
 I_3= \int \frac{dk \, d \omega_1}{[2 \pi]^{2}} \int \frac{dp \, d\omega_2}{[2 \pi]^{2}}
 [k \, p] ([k+p])^2 |G(\hat{k})|^2 |G(\hat{p})|^2 G(-\hat{k}-\hat{p}) \ln(\Lambda_0/k)
 \label{integral-frm-I3}
 \eeq
  \begin{figure}[ht]
\vskip-0.1in
\hskip.8in \includegraphics[scale=.6]{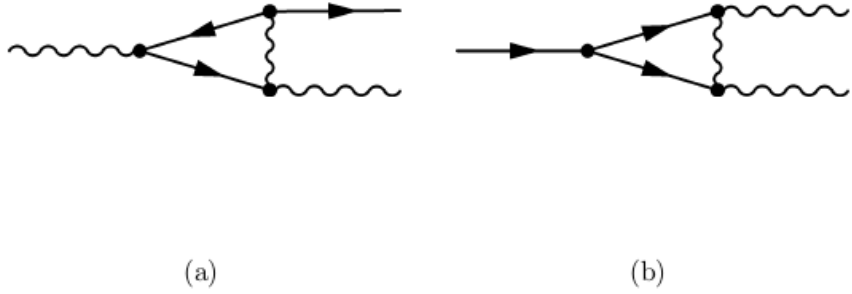}
\vskip-0.3in
\caption{Feynman diagrams: (a) $W^{(3)}_3$ and (b) $W^{(4)}_3$.
These diagrams yield logarithmic corrections same in magnitude with opposite signs.}
\label{fig2}
\end{figure}
 \subsection{Calculation of $W^{(4)}_3$}
We express $W^{(4)}_3$ in momentum and frequency spaces as  
\begin{eqnarray}
 W^{(4)}_3 &=& \left(\frac{-\lambda_0}{2}\right)^3 \int \frac{d^{d+1} \hat{k}}{[2\pi]^{d+1}} 
 \int \frac{d^{d+1} \hat{k}'}{[2\pi]^{d+1}} \int \frac{d^{d+1} \hat{p}}{[2\pi]^{d+1}} 
 \int \frac{d^{d+1} \hat{q}}{[2\pi]^{d+1}} \int \frac{d^{d+1} \hat{q}'}{[2\pi]^{d+1}}
 \int \frac{d^{d+1} \hat{Q}}{[2\pi]^{d+1}}  \nonumber \\
 && \times [\mb q \cdot (\mb k'-\mb Q-\mb q)] [\mb Q \cdot (\mb k'- \mb Q)] [\mb q' \cdot (\mb Q- \mb q')] |\mb k'-\mb Q|^2
|Q|^2 |k'|^2   G(\hat{Q}) G(\hat{k}')   \nonumber \\
&& \times  G(\hat{k}'-\hat{Q})  [\langle h(\hat{p}) h(\hat{k}) h(\hat{q}) h(\hat{q}') h(\hat{k}'-\hat{Q}-\hat{q}) h(\hat{Q}-\hat{q}') 
\rangle]
 \label{W43-ini-form}
\end{eqnarray}
and the corresponding Feynman diagram is depicted in figure\,3(b).
 Height correlation in equation \ref{W43-ini-form} is written as
 \hspace{-1cm}
 \begin{eqnarray}
 \langle h(\hat{p}) h(\hat{k}) h(\hat{q})  h(\hat{q}')   h(\hat{k}'-\hat{Q}-\hat{q}) h(\hat{Q}-\hat{q}') \rangle &=&
 24 (2D_0)^3 [2\pi]^{3(d+1)} |G_0(\hat{k}+\hat{Q})|^2 |G_0(\hat{p})|^2  \nonumber \\
 && |G_0(\hat{k})|^2  \delta^{d+1}(\hat{q}+\hat{k}+\hat{Q}) \delta^{d+1}(\hat{Q}+\hat{q}'-\hat{k}) \nonumber \\
 && \delta^{d+1}(\hat{k}+\hat{k}'+\hat{p}).
 \label{W43-6height-corr}
\end{eqnarray}
  We substitute equation \ref{W43-6height-corr} in equation \ref{W43-ini-form} and obtain 
  \begin{eqnarray}
  W^{(4)}_3 &=& -24 (\lambda_0 D_0)^3 \int \frac{d^{d+1} \hat{k}}{[2\pi]^{d+1}} \int \frac{d^{d+1} \hat{p}}{[2\pi]^{d+1}} 
  |\mb k+\mb p|^2 |G(\hat{p})|^2  |G(\hat{k})|^2] G(\hat{k}-\hat{p}) L^{(4)}_3(\hat{p}; \hat{k}) \nonumber  \\
 \label{W43-ampu-short}
\end{eqnarray}
 where 
\begin{eqnarray}
 l^{(4)}_3(\hat{p};\hat{k})&=&\int \frac{d^{d+1} \hat{Q}}{[2 \pi]^{d+1}}  [\mb Q \cdot (-\mb k-\mb p-\mb Q)] [(-\mb k-\mb Q) \cdot -\mb p ] 
  [(\mb k+\mb Q) \cdot -\mb k ] |\mb k+\mb p+\mb Q|^2  \nonumber \\
 && |Q|^2 G_0(\hat{Q})   G(-\hat{k}-\hat{p}-\hat{Q}) |G(\hat{k}+\hat{Q})|^2 
 \label{L43-bef-renorma}
\end{eqnarray}
Considering external momenta and frequencies ($\hat{p}$ and $\hat{k}$) are much smaller than the internal momenta and 
frequency ($\hat{Q}$), we obtain 
\begin{equation}
 l^{(4)}_3(\mb p,0;\mb k,0)=\int \frac{d^{d+1} \hat{Q}}{[2 \pi]^{d+1}} \, [\mb Q \cdot \mb k]  \, [\mb Q \cdot \mb p] 
 |Q|^6 \, G_0(\hat{Q})  G(-\hat{k}-\hat{p}-\hat{Q}) |G(\hat{Q})|^2 
 \label{L43-aft-approx}
\end{equation}
Performing the frequency integration and working out the momentum integration in the high-momentum shell 
$\Lambda_0 e^{-r} \leq Q \leq \Lambda_0$, we follow the same procedure as in Sec. 4.1 and 4.2 and obtain
\beq
 l^{(4)}_3(p,0;k,0)=-\frac{p \, k}{5 \lambda^2_0 D_0} \ln(\Lambda_0/k)
 \label{L43-final-form}
 \eeq
 in $d=1$ dimension. We substitute equation \ref{L43-final-form} in equation \ref{W43-6height-corr} and get
 \begin{equation}
 W^{(4)}_3 =-\frac{24}{5} \left(\lambda_0 D^2_0 \right) \, \, I_4
\label{W43-final-form}
 \end{equation}
where
\beq
I_4=\int \frac{dk \, d \omega_1}{[2 \pi]^{2}} \int \frac{dp \, d\omega_2}{[2 \pi]^{2}}
 [k \, p] ([k+p])^2 |G(\hat{k})|^2 |G(\hat{p})|^2 G(-\hat{k}-\hat{p}) \ln(\Lambda_0/k).
\label{integral-frm-I4}
 \eeq
It is seen that the integral form of $I_3$ and $I_4$, expressed in equations \ref{integral-frm-I3} and 
\ref{integral-frm-I4}, respectively, are identical. Hence the logarithmic contributions $W^{(3)}_3$ and $W^{(4)}_3$, 
obtained in equations \ref{W33-final-form} and \ref{W43-final-form}, respectively,  are same in magnitude and opposite 
in signs. Thus, they cancel each other exactly and effectively do not contribute to $W_3$.

\subsection{Calculation of $W^{(5)}_3$}
The expression for $W^{(5)}_3$ in momentum and frequency space is given by
\begin{eqnarray}
W^{(5)}_3&=& \left(\frac{-\lambda_0}{2}\right)^3  \int \frac{d^{d+1} \hat{k}}{[2\pi]^{d+1}}  
\int \frac{d^{d+1} \hat{k}'}{[2\pi]^{d+1}}\int \frac{d^{d+1} \hat{p}}{[2\pi]^{d+1}}
\int \frac{d^{d+1} \hat{q}}{[2\pi]^{d+1}} \int \frac{d^{d+1} \hat{q}'}{[2\pi]^{d+1}} 
\int \frac{d^{d+1} \hat{Q}}{[2\pi]^{d+1}} \nonumber \\
&&  [\mb Q \cdot (\mb k'- \mb Q)] [\mb q \cdot (\mb p- \mb q)] 
[\mb q' \cdot (\mb Q- \mb q')] |k'|^2 |p|^2 |Q|^2 G(\hat{k'}) G(\hat{p}) G(\hat{Q})  \nonumber \\
&&  \langle h(\hat{k}'-\hat{Q}) h(\hat{q}') h(\hat{Q}-\hat{q}') h(\hat{q})h(\hat{p}-\hat{q}) h(\hat{k}') \rangle
\label{W53-ini-form}
\end{eqnarray}
Feynman diagram corresponding to $W^{(5)}_3$ is shown in figure 4(a). We express the height correlation 
appearing in equation \ref{W53-ini-form} as
\begin{eqnarray}
\langle h(\hat{k}'-\hat{Q}) h(\hat{q}') h(\hat{Q}-\hat{q}') h(\hat{q})h(\hat{p}-\hat{q}) h(\hat{k}) \rangle&=&48 (2D_0)^3 
[2\pi]^{3(d+1)} G(\hat{p}+\hat{k}') G(\hat{k})   |G(\hat{k}')|^2 \nonumber \\
&&  |G_0(\hat{k}'-\hat{Q})|^2 \, \delta^{d+1}(\hat{k}'+\hat{q}) \, \delta^{d+1}(\hat{k}'-\hat{q}') \nonumber \\
&& \delta^{d+1}(\hat{k}+\hat{p}+\hat{k}').
\label{W53-6height-corr}
\end{eqnarray}
Substituting equation \ref{W53-6height-corr} in equation \ref{W53-ini-form}, we obtain 
\begin{eqnarray}
 W^{(5)}_3&=& 48 (2D_0)^2 \left(\frac{-\lambda_0}{2}\right)^3  \int \frac{d^{d+1} \hat{k}}{[2\pi]^{d+1}}
\int \frac{d^{d+1} \hat{p}}{[2\pi]^{d+1}} |\mb k+\mb p|^2 |p|^2 G(-\hat{k}-\hat{p}) G(\hat{p}) G(\hat{k}) \nonumber \\
&& |G(\hat{k}+\hat{p})|^2  [(\mb k+\mb p) \cdot \mb k ] L^{(5)}_3(\hat{k}+\hat{p})
 \end{eqnarray}
where 
\begin{eqnarray}
 L^{(5)}_3(\hat{k}+\hat{p})&=& (2D_0) \int \frac{d^{d+1} \hat{Q}}{[2\pi]^{d+1}} [\mb Q \cdot (\mb Q+\mb k+\mb p)] 
 [(\mb Q-\mb k-\mb p) \cdot (\mb k+\mb p)] |Q|^2 G_0(\hat{Q}) \nonumber \\
 && |G_0(\hat{k}+\hat{p}+\hat{Q})|^2
\end{eqnarray}
Considering internal momenta are much larger than the external momenta, we obtain 
\begin{equation}
L^{(5)}_3(\hat{k}+\hat{p})= -(2D_0) \int \frac{d^{d+1} \hat{Q}}{[2\pi]^{d+1}} |Q|^4 
[\mb Q \cdot (\mb k+\mb p)] G_0(\hat{Q}) |G_0(\hat{k}+\hat{p}+\hat{Q})|^2
\end{equation}
We perform the frequency integration and obtain 
\beq
L^{(5)}_3(k+p)= -\frac{D_0}{4 \nu^2_0} (k+p) \int_{-\infty}^{\infty} \frac{d Q}{[2 \pi]} \, \, \frac{Q }{|Q|^4} =0
\eeq
in $d=1$ dimension.
Hence, $W^{(5)}_3$ yields zero contribution. Similarly other terms such as $W^{(6)}_3$, $W^{(7)}_3$ and $W^{(8)}_3$, 
shown in Figs. 4(b), 4(c) and 4(d), respectively, give vanishing contributions individually.

\begin{figure}[ht]
\vskip-0.1in
\hskip.9in \includegraphics[scale=0.6]{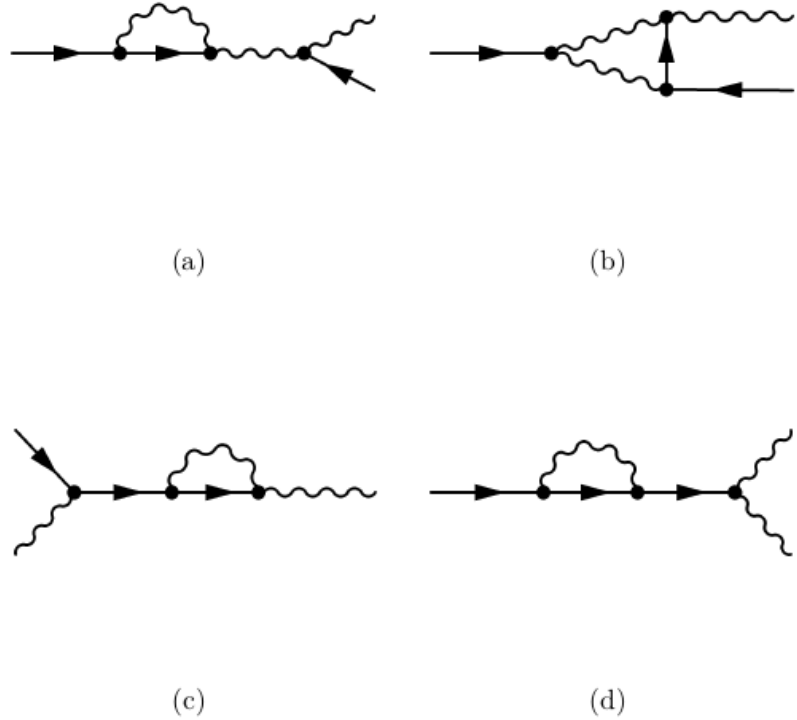}
\vskip-0.3in
\caption{Feynman diagrams: (a) $W^{(5)}_3$, (b) $W^{(6)}_3$, (c) $W^{(7)}_3$ 
and (d) $W^{(8)}_3$. These diagrams yield zero contributions individually.}
\label{fig2}
\end{figure}

\subsection{Skewness in One dimension}
Adding all the terms on the right hand side of equation \ref{petur-W3}, we observe that the effective 
contribution to $W_3$ comes from $W^{(1)}_3$ and $W^{(2)}_3$, given in equations \ref{W2-final} and \ref{W3final}, yielding   
\begin{eqnarray}
W_3=W^{(1)}_3+W^{(2)}_3=-\frac{3}{50 \pi} [0.249917] \left(\frac{D_0}{\lambda_0}\right) \frac{1}{\mu^3}.
\label{W3final}
\end{eqnarray}

Hence, we calculate the skewness employing the definition given in equation \ref{skew} wherein 
we substitute the calculated values for $W_2$ and $W_3$ from equations \ref{W2-final} and \ref{W3final} . Thus we obtain 
\beq
S=\frac{W_3}{W^{3/2}_2}=-0.044072
\eeq
which is independent of the model parameters ($\lambda_0$ $\nu_0$ $D_0$ ) and the momentum cutoffs ($\mu_0$, $\Lambda_0$).

\section{Discussion and Conclusion}
In this paper, we considered the conserved stochastic growth of a surface due to particle deposition 
on a flat substrate and we find the statistical properties of the surface in the large-scale long-time limits. 
The corresponding dynamics is represented by the VLDS equation driven by a 
stochastic Gaussian white noise. We followed an RG scheme without rescaling akin to 
Yakhot and Orszag \cite{yakhot_j_s_comput_1_3_86,yakhot_prl_57_1772_86}.  

This RG scheme differs 
from the conventional dynamical RG scheme of Ma and Mazenko \cite{s.k.Ma_prb_11_4077_75}, Forster \emph{et al.} \cite{forster_pra_16_732_77}, and 
Medina \emph{et al.} \cite{KPZ89} in the sense that the renormalized quantity are not rescaled and their flow equation 
are directly worked out. Thus, we obtained the 
renormalized diffusivity $\nu(r)$ at large $r$ and the roughness exponent $\chi=1$. This is consistent with the previous
RG results obtained by Lai and Das Sarma  \cite{VLDS_91}.

Subsequently,  we calculate the skewness of height fluctuations of the VLDS equation. 
A similar scheme was employed earlier for the calculation of skewness 
\cite{Singha-Nandy14_PRE.90.062402} and kurtosis \cite{Singha-Nandy15_jstatmech} in the case of 
non-conserved interface growth governed by the $(1+1)$-dimensional KPZ equation. 
The amputated parts of the connected loop diagrams, namely, $l_2$, $l^{(1)}_3$, $l^{(2)}_3$, $l^{(3)}_3$,   
$l^{(4)}_3$ etc., for the second and third moments as shown in Figs.\ 1(b), 2(a), 2(b), 3(a), 3(b) etc., respectively,
are renormalized by means of scale elimination and by constructing differential equations representing how these 
renormalized quantities flow with respect to the RG decimation scheme.
The solutions to the differential equations immediately yield their renormalized values.
Employing this RG scheme, we directly find the renormalized expressions for these loop integrals 
which are indispensable for the calculation of moments. These renormalized amputated parts of the loops 
are used in  conjunction with the external legs to numerically evaluate the integrals for the second and third moments. 

Our approach subsumes calculations of $W_2$ and $W_3$ in the large scale limit $k\rightarrow 0$
and consequently expects the statistical properties of the growth process at large scales. It is to be noted that 
these moments are independent of the upper cutoff $\Lambda_0$ and functions of lower cutoff $\mu$
which may be identified with the inverse of the substrate size $L$. These moments are also observed to follow the 
expected scaling $W_n \sim L^{n \chi}$. The value for the skewness
is immediately obtained from these calculated values which is found to be independent of the model parameters 
($\lambda_0$, $D_0$, $\nu_0$) and the UV and IR cutoffs ($\Lambda_0$, $\mu$), suggesting its universality.

It is interesting to note that we obtained the skewness value  
which is a negative number. This is consistent with the numerical prediction of Das Sarma \emph{et al.}
\cite{Das-Sarma_PRE53_1996} who obtained $S=-0.1\pm 0.15$ in the steady state and suggested that the skewness value 
is likely to be negative, although they did not exclude a zero or slightly positive value. 
The large error bar that they obtained is probably due to a dominant role of fluctuations 
in their numerical model. We observe that our value of $S=-0.0441$ is consistent with their
prediction. Moreover our calculated value  asserts that the probability distribution function
is negatively skewed. 

\subsection*{Acknowledgements} 
T.S. is thankful to the Ministry of Human Resource Development (MHRD), Government of India, 
for financial support through a scholarship.

\end{document}